# The current spin on manganites

C. Israel[1], M. J. Calderón[2] and N. D. Mathur[1*]

[1]Department of Materials Science, University of Cambridge, Cambridge CB2 3QZ, UK
[2]Instituto de Ciencia de Materiales de Madrid (CSIC), Cantoblanco, 28049 Madrid, Spain.
*E-mail: ndm12@cam.ac.uk

**STANDFIRST:**
Manganites are pseudo-cubic oxides of manganese that show extremes of functional behavior. Diverse magnetic and electronic phases coexist on a wide range of length scales even within single crystals. This coexistence demonstrates a complexity that inspires ever deeper study. Yet even the basic nature of the coexisting phases remains controversial. Can the ferromagnetic metallic phase provide fully spin-polarized electrons for spin electronics? Does the superlattice in the highly insulating phase represent charge order? Here we highlight recent results that demonstrate a coexistence of opinions about a field in rude health.

**ABSTRACT:**
**In a material, the existence and coexistence of phases with very different magnetic and electronic properties is both unusual and surprising. Manganites in particular capture the imagination because they demonstrate a complexity that belies their chemically single-phase nature. This complexity arises because the magnetic, electronic and crystal structures interact with one another to deliver exotic magnetic and electronic phases that coexist. This coexistence is self-organized and yet readily susceptible to external perturbations, permitting subtle and imaginative experiments of the type that we describe here. Moreover, these experiments reveal that each competing phase itself remains an incompletely solved mystery.**

Manganites were known to show pronounced magnetoresistance[1,2] and phase separation[3] effects in the 1950s, but they really hit the heights during the 1990s for three reasons. First, magnetoresistance took prominence between the 1988 discovery[4] of giant magnetoresistance (GMR) in metallic multilayers, and the first shipment of GMR disc-drive heads by IBM in 1998. Second, great advances in laboratory infrastructure permitted new approaches, e.g. the fabrication of high-quality thin films, the advent of superconducting magnets, and the imaging of magnetic and electronic texture via e.g. scanning probe techniques. Third, this infrastructure had been very productive in the study of the high-temperature cuprate superconductors, and scientists were therefore immediately able to accept the new challenges presented by the manganites.

The manganites are a family of perovskite oxides in which the composition of the A-site cations may be varied using mixtures of divalent rare-earth and trivalent alkaline-earth elements. The most immediate consequence of this variation is to alter the charge doping of the magnetic and electronic structures that reside in the sublattice formed by the B-site manganese and intervening oxygen atoms. Significantly charge disorder on the

A-sites has little effect on physical properties[5,6], suggesting that it is screened. However, the average size of the A-site cations plays a strong role in determining physical properties, as does size disorder[7]. This is because the A-site cations are relatively small such that the surrounding $MnO_6$ octahedra are tilted. Physical properties are also strongly influenced when these octahedra undergo Jahn-Teller distortions due to the presence of valence electrons. And extrinsic and intrinsic strain are so significant[8] that epitaxial films, single crystals, powders and polycrystalline samples all have the potential to behave quite differently.

A suitable mix of similarly sized cations makes it favorable for Mn valence electrons (in $3d$ $e_g$ levels) to delocalize and magnetically link Mn $3m_B$ core spins, i.e. three highly localized $3d$ $t_{2g}$ electrons that are strongly Hund coupled to each other and on-site valence electrons. Thus, for example, $La_{0.7}Ca_{0.3}MnO_3$ is a ferromagnetic metal (FMM) below a Curie temperature[9] of $T_C \sim 260$ K. Above this temperature the system is a paramagnetic insulator (PMI) because the Mn core spin directions are random, and the valence electrons are bound to lattice distortions to form polarons that are not very mobile. Near $T_C$, an applied magnetic field aligns the core spins and recovers the metallic state. This negative magnetoresistance (MR) was found to be colossal ($\sim 10^5$%) in a strained annealed film[10], the term colossal magnetoresistance (CMR) was coined, and the field of manganites was born again.

Even larger values of MR ($\sim 10^{12}$% in $Pr_{0.67}Ca_{0.33}MnO_3$) are recorded if the magnetic field is applied to the more highly insulating phases[11] of manganites that result from the use of small A-site cations such as $Pr^{3+}$. The traditional cartoon of charge order (CO) used to describe these insulating states involves planes of $Mn^{3+}$ interspersed with planes of $Mn^{4+}$, and antiferromagnetic core-spin order. As we discuss later, the appearance of a superlattice in the CO phase is unambiguous in a range of diffraction experiments, but the underlying source of the modulation is controversial. However, we will persist with the CO epithet since it is unlikely that charge modulation is completely absent in the superlattice, and in any case the label is familiar.

Certain manganites such as $(La,Pr,Ca)MnO_3$ show phase separation between the FMM and CO phases. The spatial distribution of these phases was first visualized[12] using transmission electron microscopy (TEM) to reveal the extent of the CO phase. Later, a combination of TEM probes[13] sensitive to each of the coexisting phases revealed a phase that was unexpectedly both FMM and CO. This hints at how subtly the manganites can behave, a trait reflected in the controversy regarding the CO phase itself. Many further studies have revealed phase coexistence phenomena on length scales that are so short that it is questionable whether each coexisting region truly represents a thermodynamic phase[14].

In this review we comment on a snapshot of recent manganite results, passing over the earlier discoveries that are described elsewhere[15-22], and recent work on "layered" manganites[23-25] in which rocksalt layers separate the perovskite manganite blocks. Recent review articles[26-31] to which the manganites are relevant cover oxide spin electronics (spintronics); the coexistence of magnetic and electrical order (multiferroics); and magnetoelectric coupling between magnetic and electrical order parameters. This diversity of contexts epitomizes the multifarious nature of the manganites.

**Phase separation**

Phase separation is anticipated between ordered phases (e.g. the FMM and CO phase[12,13]) that form below temperatures as high as room temperature, and atomically sharp phase boundaries have been observed in scanning tunneling microscopy (STM) experiments on a single crystal[32]. Coexistence between one of these ordered phases and the PMI, seen in thin films by STM[33,34], does not[23] occur in every manganite, and could be an extrinsic effect due to thin-film strain or methodology[35]. To address this issue, Steffen Wirth and colleagues constructed zero-bias STM conductance maps[35] for a $Pr_{0.68}Pb_{0.32}MnO_3$ single crystal. They found a nanoscale phase separation (Fig. 1) that appears to be intrinsic because (i) it only occurred at temperatures near the metal-insulator transition, and (ii) it is far smaller than the length scales associated with twinning. The phase distribution should therefore vary at least from run to run, but this was not confirmed.

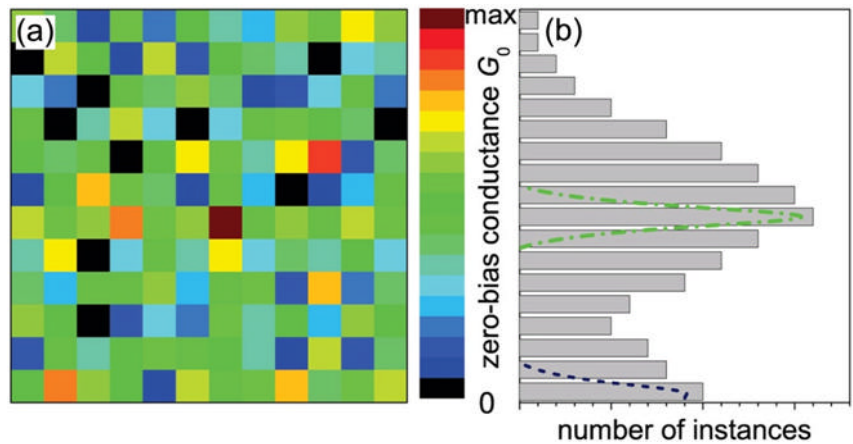

Fig. 1 (a) Schematic zero-bias STM conductance map for a $Pr_{0.68}Pb_{0.32}MnO_3$ single crystal at a temperature near the metal-insulator transition (typical pixel resolution ~1 nm). (b) The corresponding conductance values are represented on this histogram. Two peaks are apparent, reflecting a phase separation that appears to be intrinsic because at higher (lower) temperature the sample is significantly more homogeneous such that (i) the conductance map is mainly blue (green) and (ii) the histogram displays only a single blue (green) peak. Images courtesy of Steffen Wirth, based on [35].

Magnetic force microscopy (MFM) has previously revealed[36] that the FMM component in a phase-separated $(La,Pr,Ca)MnO_3$ film adopts a pattern that varies as the temperature is changed. This fluidity was somewhat surprising, since long-range strain arising from the nucleation of FMM regions had the potential to globally lock the pattern of phase separation. A more recent surprise from the same laboratory was the discovery[37] that although $(La,Pr,Ca)MnO_3$ shows glassy magnetic behavior[38], it is not a spin-cluster glass[39]: variable-field MFM studies[37] found that sample magnetization was determined by FMM phase fraction rather than cluster alignment[12]. Interestingly, this experiment recorded correlations between magnetic phase changes and twin boundaries (Fig. 2), consistent with the influence of the crystal structure, and strain, on physical properties.

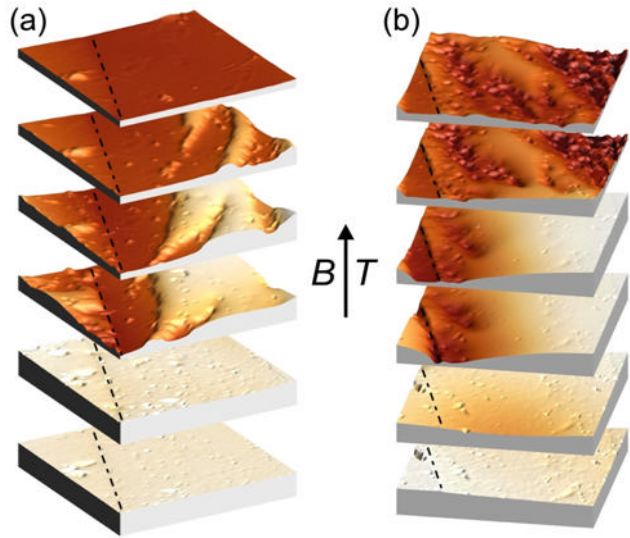

**Fig. 2 MFM images of a (La,Pr,Ca)MnO$_3$ single crystal showing the conversion of the CO phase (bright) to the FMM phase (dark) under (a) the influence of an increasing magnetic field $B$ at 7 K, and (b) increasing temperature $T$ at 1 T. Twin boundaries (black dashed lines) serve as both nucleation sites and boundaries for phase conversion. Images are (a) 7×7 μm$^2$ and (b) 6×6 μm$^2$. After [37].**

TEM has been very valuable in the study of manganites over the last decade[12,13], and continues to produce important results. Recently, polycrystalline (La,Ca)MnO$_3$ that nominally assumes an orthorhombic CO phase was found to contain micron-sized monoclinic needle twins with no CO modulation[40,41]. It is instructive to compare with the afore-mentioned STM studies[35] that report intrinsic phase separation in single crystal Pr$_{0.68}$Pb$_{0.32}$MnO$_3$. In polycrystalline (La,Ca)MnO$_3$, the monoclinic phase appears to be extrinsic since x-ray diffraction reveals it to be absent in powders that are nominally strain free[42].

Phase coexistence in manganites can produce distinctive electrical and magnetic memory effects[43]. The changes observed under the influence of an applied magnetic field superficially resemble the sharp switches associated with the reorientation of FMM domains at discontinuities such as grain boundaries[44-46] and tunnel barriers[47-49]. However, just like the original CMR effect[10], memory effects[43] in phase-separated manganites involve such a pronounced continuous response that even small magnetic fields result in a significant change. And if these small magnetic fields are turned on quickly, the responses they generate can appear sharp on a suitable timescale. Electric rather than magnetic fields can produce switching in various transition metal oxides due to the influence near current contacts of charge injection/extraction on sample oxygenation. A pronounced electrical pulse-induced resistance-change (EPIR) has been seen in manganites that are prone to phase separation[50,51], but also those that are not[52,53]. Local[53] switching and imaging seems a particularly promising route to probe further the nature of this distinction.

**Phase separation vis-à-vis CMR**
These beautiful recent results demonstrating phase separation build on the earlier excitement regarding CMR. It is therefore interesting to consider not just why phase separation arises in the manganites, but how it relates to CMR.

Phase separation may be considered to arise in connection with the complexity that is reflected in the manganite Hamiltonian, which should include:
- the kinetic energy of the conduction electrons plus Hund coupling (double exchange[54]),
- a direct antiferromagnetic coupling between the core spins,
- electron-phonon coupling between octahedral deformations and the $e_g$ orbitals,
- on-site and long-range Coulomb interactions, and
- diagonal (on-site) and off-diagonal (hopping) disorder.

Given this complexity, it is not surprising that the energy landscape has several local minima lying at similar energy levels. Each minimum corresponds to a different phase, and experimentally it is easy to take the system from one minimum to another using various control parameters such as temperature.

Extrinsic phase separation is commonly observed near either first- or second-order transitions as a consequence of the local energetics being modified by defects or strain, i.e. nucleating agents that render phase distributions similar from run to run. Even in the absence of such agents, intrinsic phase separation near first-order transitions is expected in principle[55] and arises in practice unless kinetically prevented; the new phase is nucleated by intrinsic features such as magnetic fluctuations, and the phase distribution varies from run to run. Although phase separation is pronounced in the manganites, it does not require all the ingredients of manganite physics, and can indeed arise from a simple double-exchange model[55], or strain in a model where the electronic and magnetic degrees of freedom are not taken into account[56]. But the richness of manganite physics is significant because it sometimes produces phase separation on such a short length scale that the result is a new thermodynamic phase displaying magnetic and electronic texture[14].

Although it is believed that phase separation typically underpins CMR[57], and it is true that phase separation strongly influences CMR, phase separation is not in fact formally required for CMR. A CMR response arises when a magnetic field interconverts two phases which are very different, and thus separated by a first-order boundary near which phase separation is expected. However, in principle, a CMR response could arise in a homogenous system, and indeed the layered manganite $La_{1.4}Sr_{1.6}Mn_2O_7$ shows a unimodal distribution of zero-bias conductances, indicating no phase separation[23]. Therefore the role of phase separation in CMR is circumstantial and not causal. That said, phase separation will strongly influence e.g. current pathways and therefore CMR.

**Devices**
Electronic and magnetic devices are of course widely used to great effect in an industrial context, e.g. transistors in computer chips, and magnetic sensors for anti-lock brakes in automobiles. Thin-film devices also represent scientific tools for manipulating and monitoring the manganites. For example, lithographic patterning on some length scale can indicate the presence[58,59] or absence[60] of phase separation on that length scale. In another example, devices designed to trap magnetic domain walls in the FMM phase[61,62] hint at current-induced domain wall deformation[63], but clear evidence for the proposed[14] phase separation at wall centres remains elusive.

Magnetic tunnel junctions made from traditional magnetic metals are currently starting to find their way into data storage applications[64]. Some of the first manganite devices were epitaxial tunnel junctions[47] made from FMM electrodes of (La,Sr)MnO$_3$ or (La,Ca)MnO$_3$ with an ultra-thin SrTiO$_3$ barrier. Subsequent improvements[48,49] were both qualitative and quantitative, producing reports of clean low-field MR switching with a magnitude on the order of 1000%. Recent improvements in performance (Fig. 3) are based on reducing interfacial charge discontinuities[65] by interface engineering[66]. This approach could be relevant to following through on the interesting suggestion[24] that a self-organized barrier forms at the free surface of a layered manganite. Separately, it would be interesting to fully explore in manganite tunnel junctions the role of symmetry filtering[26,27,67] by crystalline barriers, a phenomenon that has been very successfully exploited to produce large room-temperature MR using MgO barriers with CoFe[68] or Fe[69] electrodes.

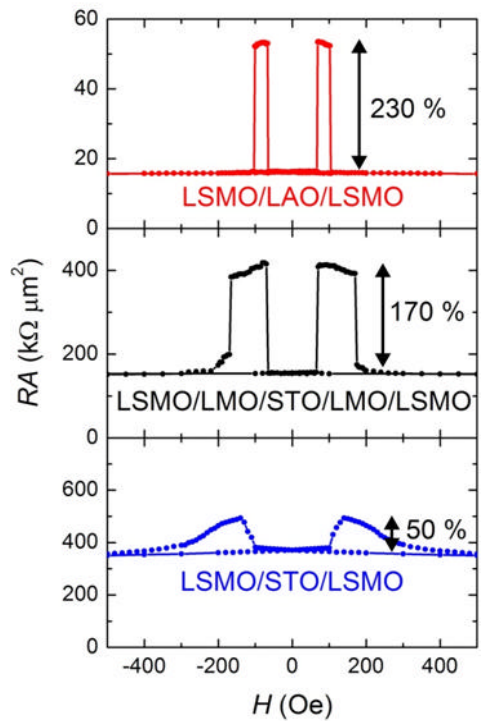

**Fig. 3 MR data at 10 K for magnetic tunnel junctions with (La,Sr)MnO$_3$ (LSMO) electrodes. An SrTiO$_3$ (STO) barrier (bottom panel) was used for the original manganite tunnel junctions[47]. Replacing it with a trilayer that also involves LaMnO$_3$ (LMO) is designed to increase the MR (middle panel) by compensating interfacial charge transfer. Using instead an LaAlO$_3$ (LAO) barrier is designed to prevent charge transfer, and this produces an even larger MR (top panel). After [66].**

MR devices may also be obtained without explicitly incorporating a tunnel barrier by growing a bilayer comprising a manganite and some other FMM material such as magnetite[70] or permalloy[71]. Magnetic decoupling is thought to be the result of interfacial effects associated with e.g. structural discontinuities or oxygenation. Alternatively, MR devices may be obtained using FMM manganite electrodes separated over relatively large distances by organic materials. This is because carbon has a low atomic number and therefore weak spin-orbit coupling, which means that traveling spin-polarized electrons do not lose their magnetic information. Sub-micron organic layers of 8-hydroxy-quinoline aluminum (Alq3) between (La,Sr)MnO$_3$ and Co electrodes are the basis for devices[72] that show a large low-field MR~30%. Moreover, hysteretic current-voltage characteristics in similar devices[73] produce memory effects reminiscent of those due to electric fields at manganite contacts[51].

Carbon nanotubes are not taxonomically considered to belong to the organic family, but exploiting their high Fermi velocity[74] ($0.8 \times 10^6$ m s$^{-1}$) as well as weak spin-orbit coupling permits the transport of spin information over micron-scale distances between (La,Sr)MnO$_3$ electrodes[75]. The associated conversion of magnetic information into large electrical signals corresponds to an MR of 61% (Fig. 4), and represents the basis for a spin transistor if the nanotube can be suitably gated. The innate advantage of a spin transistor is that the use of a magnetic gate would permit non-volatile information storage, unlike today's fast silicon transistors. And the surprising success with nanotubes and manganites is particularly significant[75] given that the relevant figure of merit for spin transistor-like devices based on semiconductors[76] is an MR of only ~1%. Note that nanotubes *of* manganites[77,78] (Fig. 5) have yet to be incorporated in spintronic devices.

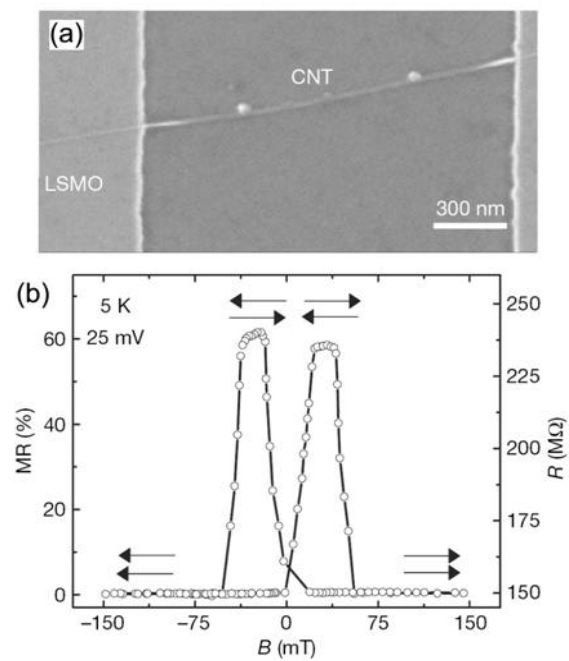

**Fig. 4 (a) Scanning electron microscopy image of a 20 nm diameter multiwall carbon nanotube (CNT) running between epitaxial La$_{0.7}$Sr$_{0.3}$MnO$_3$ (LSMO) electrodes. (b) The MR of this two-terminal device at 5 K under 25 mV bias, with the arrows showing the magnetic orientations of the**

manganite electrodes. The large MR persists up to a higher bias of 110 mV, permitting magnetic information to be converted to large electrical signals. CNTs could therefore make an impact in spintronics, e.g. as the non-magnetic channel in spin transistors. After [75].

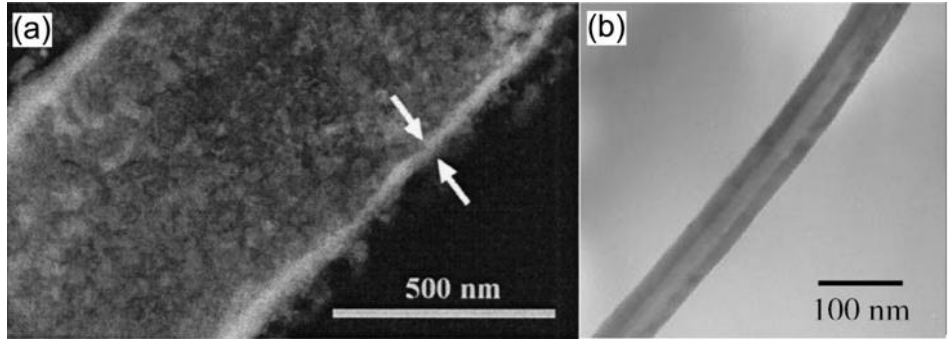

**Fig. 5** TEM images of (a) a $(La,Sr)MnO_3$ nanotube synthesized via a pore-wetting process with liquid precursors[77], and (b) a $MgO-(La,Sr)MnO_3$ core-shell nanowire fabricated by the pulsed laser deposition of $(La,Sr)MnO_3$ onto single-crystalline MgO nanowires[78].

The devices discussed above require the magnetic electrodes to be switched using an applied magnetic field. For commercial applications using traditional magnetic metals, there is currently widespread interest in magnetic switching using the torque delivered by spin-polarized currents[79]. Demonstrating the efficacy of using manganites as a test-bed for novel effects, Jonathan Sun of IBM (Yorktown Heights) first reported[80] spin torque in a manganite structure via the observation of asymmetric current-voltage characteristics. The magnetically switched object was an intergrowth between FMM manganite electrodes in a device similar to the tunnel junctions he pioneered[47]. The architecture employed for this report of spin torque is somewhat ironic given that defects in tunnel barriers are normally associated with device failure. But, asymmetric current-voltage curves attributed to spin torque have now[81] been reported in planar nanoconstrictions[61-63] in a $(La,Ba)MnO_3$ film.

The related effect of switching a magnetization using an electric field, rather than a current, is currently a popular goal in manganite device studies. This is primarily because it mimics the attractive electric-write process in ferroelectric random access memory (FeRAM) without mimicking the less attractive magnetic-write processes associated with magnetic MRAM. One effective strategy is to apply a voltage between the surface of a manganite film and the underside of an underlying ferroelectric substrate, in order to exploit strain coupling at the interface. This strategy requires the ferroelectric to also be ferroelastic so that it will deform when the ferroelectric domains are electrically switched. In the well-known relaxor ferroelectric perovskite $Pb(Mg,Nb)O_3-PbTiO_3$ (PMN-PT), A-site cationic disorder reduces the ferroelectric domain size to the nanoscale. PMN-PT can generate, in epitaxial manganite films, weakly hysteretic changes in magnetization[82] (Fig. 6a) that are also seen in electrical resistivity[83,84]. To achieve electrically driven switching in the resistivity[85] and magnetization[86] that is sharp and non-volatile, the relaxor should be replaced with the more traditional ferroelectric $BaTiO_3$ (Fig. 6b).

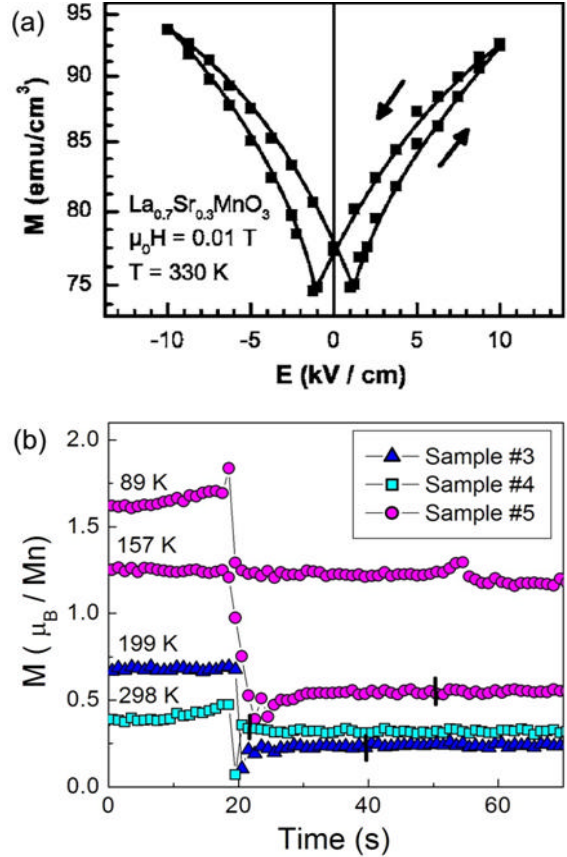

**Fig. 6** Magnetoelectric strain-mediated coupling between a ferromagnetic manganite film and its ferroelectric substrate. (a) Weak hysteresis in magnetization $M$ versus electric field $E$ applied across a (La,Sr)MnO$_3$ film and relaxor Pb(Mg,Nb)O$_3$-PbTiO$_3$ substrate[82]. (b) Sharp non-volatile switching in $M$ for three samples across which $E$ was ramped until the transition was observed[86]. Each sample comprised a (La,Sr)MnO$_3$ film on a BaTiO$_3$ substrate. The black lines indicate when $E$ was switched off.

The heterostructures discussed above[82-86] are multiferroic insofar as three ferroic orders are present between the film and the substrate, and we class[86] the electrically induced magnetic changes they display as "converse" magnetoelectric effects. For completeness, we note that "direct" magnetoelectric effects, i.e. magnetically induced changes in an electrical polarization, could in principle be used for room-temperature magnetic field sensors[87] to replace low-temperature SQUIDs. Strong interest in direct magnetoelectric effects in the manganites remains purely scientific because it is a low-temperature phenomenon, and the ferroelectric polarizations (e.g. 0.08 µC cm$^{-2}$ in[88] multiferroic TbMnO$_3$) are typically three orders of magnitude smaller than the best ferroelectrics.

Recently the themes of tunnel junctions and multiferroics were combined[89] in an all-manganite spin-filter device that displays four-state resistance behavior. Charge carriers that enter the device from an upper gold contact are spin polarized as they tunnel through an ultra-thin insulating ferromagnetic layer of (La,Bi)MnO$_3$. The magnetizations of this spin filter and an underlying FMM (La,Sr)MnO$_3$ analyzer may be switched independently by an applied magnetic field, yielding two states of resistance. The spin-filter device is therefore comparable to the tunnel junctions discussed above, but with the magnetic spin-filter layer replacing both the barrier and the upper FMM electrode of the tunnel junction. What makes the present device[89] novel is that the spin filter is multiferroic, i.e. not just ferromagnetic but also (excitingly for a manganite) robustly ferroelectric. Since the ferroelectric spin filter lies between dissimilar materials, device resistance depends on its polarization[90]. Consequently, the two magnetically encoded states become four (Fig. 7). Interestingly, exploiting this electroresistance phenomenon[89] is simpler than the FeRAM method of reading the ferroelectric polarization by cycling a voltage.

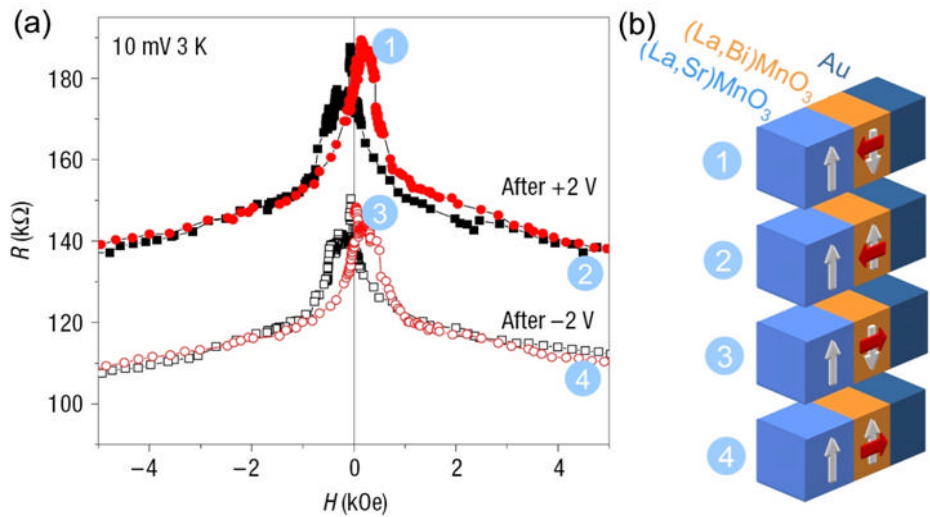

**Fig. 7 (a) For a (La,Sr)MnO$_3$/(La,Bi)MnO$_3$ (2 nm)/Au spin-filter device[89], each MR trace displays states of resistance *R* that are high (at the peaks) and low (away from the peaks). As indicated on the figure, the electrical history of the device changes the absolute resistance, thus yielding a total of four states. (b) Device schematic showing the magnetizations (white arrows) and electrical polarizations (red arrows) for the four states.**

This section has served to indicate that many manganite device applications have been suggested based on the diverse properties of the manganites. In other examples, manganite diodes[91-93] have been improved by integrating a single manganite layer with a ferroelectric and a cuprate[94] (Fig. 8), a MR>1000% has been achieved by exploiting the proximity effect in manganite-cuprate trilayers[95], and polycrystalline MR sensors have been exploited in an electromagnetic device that launches projectiles[96]. Given this diversity of performance, the continuing scientific endeavors skimmed over here justify the continued study of manganite devices even if they never find commercial success.

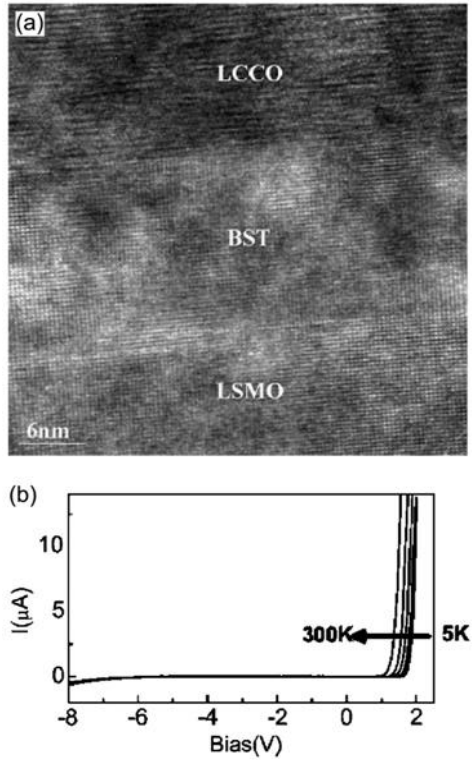

**Fig. 8** (a) High-resolution TEM cross-sectional image of a heterostructure comprising a manganite (hole-doped LSMO, $La_{0.67}Sr_{0.33}MnO_3$), a ferroelectric (BST, $Ba_{0.7}Sr_{0.3}TiO_3$), and a cuprate (electron-doped LCCO, $La_{1.89}Ce_{0.11}CuO_4$). (b) Current-voltage curves show excellent rectification over a wide range of temperatures. After [94].

**Faults in charge order**

The beguiling subtlety of the manganites is clear at time of writing, but it was not always so. In the highly insulating CO phase, the superlattice periodicity recorded in electron[97], neutron[98,99] and x-ray[98] diffraction experiments has traditionally been associated with Mn $3d\ e_g$ valence electrons residing on only those Mn atoms that lie in certain pseudo-cubic (110) planes. This simple cartoon was strongly supported by TEM cross-sectional observations[97] of "stripes", but the interpretation of such image contrast is non-trivial because the electrons from the TEM beam interact strongly with the sample causing multiple scattering (dynamical diffraction)[100]. Moreover, dark-field images showing nearly commensurate order[101] (where the periodicity of the superlattice is only just greater than double the periodicity of the parent lattice) are liable[100,102] to show interference fringes that do not represent real phenomena such as discommensurations (where the periodicity changes from one region to another).

Various re-interpretations involve the Wigner crystallization of charge[103,104], or distortions that are primarily structural[105] in which the pairing of Mn atoms has been argued[106]. However, controversy persists given that even the most sophisticated tools cannot probe the CO state with the precision required for unambiguous interpretation. This is apparent from the ability to fit neutron powder diffraction patterns equally well to more than one of the competing models[107]. Periodicity information in diffraction studies remains a far safer animal. A local TEM probe demonstrates in (La,Ca)MnO$_3$ that the superlattice possesses a period that remains uniform below a key length scale such that it cannot[100] be explained in terms of two species of pseudo-cubic (110) planes, e.g. with different Mn valences.

The complete separation of integer electronic charge across even one unit cell (to yield Mn$^{3+}$ and Mn$^{4+}$) is coulombically very expensive. Given this, and the above experimental evidence, it has now been established that the original CO picture should be replaced with a charge modulation that varies by only a small percentage from the nominal doping[108-110]. The superlattices observed in CO phases could even arise in the absence of any charge modulation due to orbital ordering, where broken symmetry due to antiferromagnetic order[3] opens a gap and stabilizes lattice distortions.

Currently, an open debate rages on whether whatever charge modulation is present in the CO phases is strongly coupled to the lattice or not. The original strong-coupling picture is based on an electron-phonon interaction that produces regions with a specific type of commensurate order (where the superlattice period is an integer multiple of the parent lattice period) separated by discommensurations[97,111]. Tight-binding models always give a charge/orbital modulation that is tied to the lattice, but the superlattice periods can be so large[110] that they cannot be experimentally distinguished from non-integer modulations. Alternatively a charge-density-wave scenario[100,112,113] with weaker (but non-zero) coupling may be more realistic. Ginzburg-Landau theory supports this picture through a complex scenario in which new thermodynamic phases[114] permit incommensurability, even at commensurate doping, due to the presence of ferromagnetism.

**Added value**
Why have manganite devices not enjoyed commercial success? There are several well-known potential factors connected with low Curie temperatures, temperature-dependent responses, high electrical resistivities and difficulties of integration with silicon technology. An extrinsic but instructive obstacle is that manganite surfaces[115] and interfaces[116-118] behave very differently from the bulk; for example, at a free surface of the FMM phase, (i) the spin polarization falls fast with increasing temperature[115] due to weak double exchange at surfaces[119], and (ii) a ferrodistortive behavior has recently been predicted[120]. This surface frailty is a manifestation of the subtle competing interactions listed earlier, and yet it is these interactions that generate the richness.

This richness inspires a question: can we use a single complex material to define a complete set of technological building blocks that mirror the vast array of single-purpose materials that we have at our disposal today? These blocks could be defined using scanning probe lithography[60] in a top-down process, but it would be more radical to control nanoscale phase separation in a manganite film without removing or adding material[121]. A more extreme phase change (between crystalline and amorphous phases) is commercially exploited in chalcogenides for CDs, DVDs, and the up-and-coming[122] all-solid-state phase-change memory (PC-RAM). In the manganites, the changes between metallic and insulating phases involve only small structural changes and far less energy.

Manganites demonstrate the potential of complex materials for device applications, and if we are ever to go beyond silicon we require basic research in as many diverse fields as possible. The manganites are excellent catalysts for interdisciplinary activity as they continue to encroach on other areas of research. For example, ferroelectricity, referred to earlier, has fallen prey to their advances and various candidate mechanisms can produce an electrical polarization[123-126]. The recent demonstration[127] that $La_{2/3}Ca_{1/3}MnO_3$ possesses a negative refractive index at GHz frequencies shows that the manganites are full of surprises. In this respect the manganites are excellent drivers to advance knowledge in a range of directions with unknown future consequences.

**Acknowledgements**
For helpful discussions we thank J Paul Attfield, Manuel Bibes, Luis Brey, Alex de Lozanne, Thomas Fix, Vincent Garcia, Leticia Granja, Luis Hueso, Harold Hwang, Pablo Levy, Peter Littlewood, Juan Salafranca, Steffen Wirth and Weida Wu. C. I. is supported by the EU FP6 STRP CoMePhS. M. J. C. acknowledges support from MAT2006-03741 and Programa Ramón y Cajal (MEC, Spain).